\documentclass[aip,pop,sd,reprint,10pt,amsmath,amssymb,graphicx]{revtex4-1}

\usepackage{bm}
\usepackage{dcolumn}
\usepackage{graphicx}
\usepackage{color}
\usepackage{caption}
\usepackage{subcaption}
\usepackage{soul}
\usepackage{siunitx}
\usepackage{verbatim}
\usepackage{float}
\usepackage{amsmath}
\usepackage{mathtools}
\usepackage{amssymb}

\newcommand{\nc}{\newcommand}
\nc{\rcite}[1]{Ref.~\onlinecite{#1}}
\nc{\rcites}[1]{Refs.~\onlinecite{#1}}
\nc{\eqeqref}[1]{Eq.~\eqref{eq:#1}}
\nc{\eqseqref}[2]{Eqs.~\eqref{eq:#1}-\eqref{eq:#2} }
\nc{\secref}[1]{Sec.~\ref{sec:#1}}
\nc{\secsref}[2]{Sec.~\ref{sec:#1}-Sec.~\ref{sec:#2}}
\nc{\ssecref}[1]{Sec.~\ref{ssec:#1}}
\nc{\ssecsref}[2]{Sec.~\ref{ssec:#1}-Sec.~\ref{ssec:#2}}

\begin{document}
\title{Comparison of edge turbulence characteristics between DIII-D and C-Mod simulations with XGC1}
\author{I. Keramidas Charidakos}
\thanks{Current address: Los Alamos National Laboratory, Los Alamos, New Mexico 87545, USA}
\affiliation{Department of Physics, University of Colorado, Boulder, Colorado 80309, USA}
\author{J. R. Myra}
\affiliation{Lodestar Research Corporation, 2400 Central Avenue, Boulder, Colorado 80301, USA}
\author{S. Ku}
\affiliation{Princeton Plasma Physics Laboratory, Princeton University, Princeton, NJ 08543-451, USA}
\author{R.M. Churchill}
\affiliation{Princeton Plasma Physics Laboratory, Princeton University, Princeton, NJ 08543-451, USA}
\author{R. Hager}
\affiliation{Princeton Plasma Physics Laboratory, Princeton University, Princeton, NJ 08543-451, USA}
\author{C.S. Chang}
\affiliation{Princeton Plasma Physics Laboratory, Princeton University, Princeton, NJ 08543-451, USA}
\author{S. Parker}
\affiliation{Department of Physics, University of Colorado, Boulder, Colorado 80309, USA}

\begin{abstract}

The physical processes taking place at the separatrix and scrape-off layer regions are crucial for the operation of tokamaks as they govern the interaction of hot plasma with the vessel walls. Numerical modeling of the edge with state-of-the-art codes attempts to elucidate the complex interactions between neoclassical drifts, turbulence, poloidal and parallel flows that control the physical set-up of the SOL region. Here, we present post-processing analysis of simulation results from the gyrokinetic code XGC1, comparing and contrasting edge turbulence characteristics from a simulation of the DIII-D tokamak against a simulation of the Alcator C-Mod tokamak. We find that the equilibrium $E \times B$ flux across the separatrix has a similar poloidal pattern in both discharges which can be explained by $\nabla B$-drifts and trapped ion excursions. However, collisionality is noted to play a major role in the way that it prevents local charge accumulations from having more global effects in the C-Mod case. In both cases, turbulent electron heat flux is observed to be higher than the ion one. This seems to be a universal characteristic of the tokamak edge, possibly related to the need of electrons to maintain quasineutrality through the only channel available to them for exiting the confinement. By Fourier analysis, we identify turbulent frequencies and growth rates of the dominant mode in both simulations. In the case of C-Mod, these numbers point to the presence of a drift wave. In the DIII-D case, further linear simulations with the {\scshape Gene} code reveal a trapped electron mode. Furthermore, using a blob detection and tracking tool, we present the amplitude and size distributions of the blobs from both simulations. The amplitude distributions are in qualitative agreement with experimental observations while the size distributions are consistent with the fact that most of the blobs are not connecting to the divertor plates and suggest that they are generated by the shearing of the turbulent modes. 
\end{abstract}
\maketitle

\section{INTRODUCTION}

A critical topic in contemporary tokamak research is the physics of the separatrix region where plasma particles and heat escape into the scrape-off layer (SOL). From the SOL, plasma impacts the divertor target plates and the main chamber walls of the device. In this region, many competitive processes are thought to be important. Poloidal and parallel flows carry plasma exhaust to the divertor, in competition with cross-field transport, both from neoclassical drifts and plasma turbulence. Some combination of these and other processes establishes the poloidal and radial structure of the SOL plasma, and the interaction of that plasma with various material surfaces.  Controlling heat fluxes on those surfaces is critical for avoiding material damage; energy distributions and particle fluxes control many interactions of interest for fusion device performance and sustainability including erosion rates, impurity sputtering and neutral recycling. For all of these reasons, it is of great importance to develop a deeper understanding of edge and SOL plasma characteristics from experimental data, theory and numerical modeling.

Qualitatively, the SOL plasma may be divided into near and far SOL regions; frequently two scale lengths of profiles are observed. \cite{LaBombard} The near-SOL region is typically characterized by steep gradients, particularly in the heat flux channel, and is of primary importance for understanding heat flow to the divertor and plasma-material interactions at the divertor target. The near-SOL heat flux width has been the subject of many experimental, theoretical and computational studies. Empirical scaling laws have been developed from a large international multi-machine database for both diverted \cite{eich2013scaling} and inner-wall limited \cite{Horacek} discharges. Fluid turbulence simulations \cite{Halpern2017} have described many features of the observed SOL gradient in the inner-wall limited case. The dominant dependence of the width with poloidal magnetic field \( B_p\) in diverted attached plasmas \cite{eich2013scaling} was modeled \cite{Pankin} and described heuristically based on neoclassical drifts \cite{Goldston2012heuristic}.  Large-scale computational efforts were undertaken to simulate the heat flux width and its scaling in both fluid \cite{ChenXu} and kinetic \cite{chang2017gyrokinetic} models. Of particular interest is the projection of the heat flux width to ITER where the scaling at high poloidal magnetic field  \( B_p\) becomes critical. This has motivated new high-field experimental results \cite{Brunner} and recent modeling efforts. A topic of considerable interest is the interplay and competition of neoclassical drift physics and turbulence. \cite{chang2017gyrokinetic}, \cite{ChenXu}, \cite{galassi2017drive}, \cite{keramidas2018analysis} If, as it is argued, \cite{Goldston2012heuristic} neoclassical drift physics, with its characteristic poloidal Larmor radius width, dominates the Eich scaling for present devices, then as  \( B_p\) increases and that width narrows, the question arises as to whether a potentially larger turbulence-controlled width will lead to a broader SOL heat flux width. Although we do not address that question specifically in this paper, the interplay of drift-orbit mechanisms and turbulence is a motivating theme of our paper.

Turning to the far SOL, convective transport by coherent turbulent structures known as blob-filaments, (also simply as “blobs” or “filaments”) is thought to be important. Observation of coherent structures in magnetic confinement devices has a long history dating back to observations on the Caltech Research Tokamak. \cite{Zweben1985} As reviewed in \rcite{Krasheninnikov2008review} and \rcite{d2011convective} blob-filaments, observed in essentially all modern fusion research devices, carry particles, energy and momentum into and through the SOL. Blob convection results from the ExB radial drift of these structures, driven for example by interchange forces which charge polarize the blob.\cite{Krasheninnikov2001}  Of interest for many applications are the net fluxes of particles and energy that are ultimately transported to the chamber walls, and the resulting plasma profiles in the SOL. SOL profiles, fluxes, blobs and statistical properties of SOL turbulence  have been studied computationally, mostly with fluid models \cite{Nespoli}, \cite{Rasmussen}, \cite{Tamain}, \cite{CohenBOUT}, \cite{Francisquez}, \cite{DudsonHERMES}, \cite{EasySTORM}, \cite{Bisai2005}, \cite{russell2015modeling}.

In order to understand and predict the net fluxes driven by blobs in the SOL, the scaling of the radial blob velocity with blob size, background plasma and magnetic field parameters has received considerable attention. Theoretical predictions of this scaling have been made \cite{Krasheninnikov2008review},\cite{d2011convective},\cite{myra2006collisionality},\cite{Manz},\cite{Wiesenberger} and tested against experimental \cite{ZwebenBlobDatabase}, \cite{Tsui}, \cite{FuchertBirkenmeier}, \cite{Higgins} , \cite{Carralero} and computational results.\cite{Paruta}, \cite{Kube}, \cite{Myra2Region}, \cite{Russell2Region} Less well understood, but under active investigation are the formation mechanisms of blobs \cite{Zhang}, \cite{Bisai}, their generation rates \cite{Fuchert}, \cite{Hacker} and their size distributions. \cite{RussellGPI} Once these quantities are determined and understood, the SOL fluxes and profiles may be modeled using statistical methods. \cite{GarciaStochastic}, \cite{militello2018two}  The blob size distribution and its relation to underlying edge and SOL instabilities are topics that will be addressed in this paper.

Our paper presents a post-processing analysis of two simulations carried out with the electrostatic version of the gyrokinetic XGC1 particle-in-cell (PIC) simulation code.  Both simulations were first used in \rcite{chang2017gyrokinetic} as part of a SOL heat-flux width scaling study.  The first one is of a DIII-D discharge that formed the basis of an earlier post-processing analysis.\cite{keramidas2018analysis} In that paper, we discussed the development of the electrostatic potential and particle flows near the separatrix that were set up in large part by the neoclassical drift orbit excursions of ions. This mechanism was shown to control the poloidal profiles of the equilibrium \(E\times B\) flows and fluxes. We showed that ambipolarity in the presence of ion orbit loss was maintained by the turbulent loss of electrons across the separatrix, and that the sheared flows set up by the ion orbit loss was of sufficient strength to affect the turbulence and impact the poloidal profile of the turbulent fluxes.

In the present paper we we extend these results, comparing this DIII-D simulation with a C-Mod simulation of higher collisionality.  Our paper has three main goals. One is to inquire into the generality of the DIII-D results described in the preceding paragraph for a higher collisionality, higher B-field device with respect to the relationship of drift-orbit and turbulence effects, sheared flows, and the poloidal profiles of the equilibrium and turbulent fluxes.  A second goal is to extend the previous work by studying the linear stability properties of the plasma in the vicinity of the separatrix.  For the DIII-D case, where an analysis with the {\scshape Gene} code\cite{jenko2000electron,gorler2010multiscale} is possible due to the fact that the dominant mode is localized within closed field lines, we identify the frequency, wavenumber and driving gradients for the most unstable modes. The final goal of our paper is to investigate the properties of blobs in these simulations, in particular their amplitude and size distribution, and to attempt to relate these observations to the instabilities and to theoretical results from blob theory. This work extends earlier studies of blobs in XGC1.\cite{churchill2017pedestal}

The plan of our paper is as follows.  In Sec.~\ref{sec:simulation} we describe and contrast the two simulation discharges under consideration. In Sec.~\ref{sec:fluxes} we examine flux patterns across the separatrix for these two cases. Linear properties of these cases are investigated in Sec.~\ref{sec:linear}, and in Sec.~\ref{sec:blobs} we consider some properties of the blob-filaments that result from these instabilities. Conclusions are given in Sec.~\ref{sec:concl}.

\section{Simulations}\label{sec:simulation}

In this section we provide a description of the two simulations domains and parameters. The simulations are of the neutral beam heated DIII-D\cite{luxon2002design} discharge \# 144981 and the RF heated Alcator C-Mod discharge \# 1100223023, with both being H-modes. They are initialized with experimental profiles taken at times $\SI{3175}  {\milli\second}$ and $\SI{1236}  {\milli\second}$ respectively. The simulation inputs include experimental profiles of electron density and temperature ($n_e$ and $T_e$), ion temperature ($T_i$), and magnetic equilibrium, from kinetic EFIT magnetic reconstructions. The geometry in both cases is lower single null but with a secondary ``virtual'' upper X-point (outside of the simulation domain). The magnetic field is in the negative $\hat{\zeta}$ direction, in cylindrical $(R, \zeta, Z)$ coordinates, making the ions $\nabla B$-drift towards the lower X-point and the electrons towards the top. In the rest of the paper, when we refer to the poloidal angle $\theta$, of a point on the separatrix, we will mean the angle in the $R-Z$ plane of a point measured from the midplane with the magnetic axis $(R_o,Z_o)$ taken as the center, i.e. $\theta = arctan\left(\frac{Z-Z_o}{R-R_o}\right)$ . Positive angles are in the counter-clockwise direction. 

In Table.~\ref{table:parameters}, we give some of the physical parameters of the two discharges. Some of them were given as input in the simulations and some others were directly calculated from the output data. More specifically, we denote with $t$ the total time of the simulation, $I_p$ is the plasma current, $B_{tor}$ is the toroidal magnetic field strength at the magnetic axis, $B_{\theta}$ is the poloidal magnetic field measured at the outboard midplane separatrix (OMP), $n_{OMP}$, $T^{OMP}_{i}$, $T^{OMP}_{e}$ are the quasineutral density of the two species, ion and electron temperature respectively, measured at OMP , $n_G$ is the Greenwald density, $\rho_i$ is the ion Larmor radius, $C_s$ is the sound speed at OMP, $V_{\theta}$ is the poloidal flow at OMP, $u_{ti}$ and $u_{te}$ are the thermal velocities of ions and electrons at OMP, $q_{95}$ is the safety factor and $qR$ the connection length measured at the surface $\Psi_N = 0.95$, $\epsilon = \frac{a}{R}$ is the aspect ratio, $\omega_{transit} = \frac{u_{te}}{qR}$ is the transit frequency of electrons, $f_{trapped}$ is the trapped particle fraction at OMP, $\tau_{ii}$ and $\tau_{ei}$ are the ion-ion and electron-ion collision times and $\lambda^{i}_{mfp}$ and $\lambda^{e}_{mfp}$ are the ion and electron mean free paths, all evaluated at OMP.  

\begin{table}
\begin{tabular}{|c|c|c|c|}
\hline
\multicolumn{3}{|c|}{Simulation Parameters}\\ \hline
    & DIII-D & C-Mod \\ \hline
 $t\;(\si{\milli\second})$ & 0.16 & 0.085 \\ \hline
 $dt\;(\si{\second})$ & 2.3 $\times 10^{-7}$ & 9.454 $\times 10^{-8}$ \\ \hline
 $I_p\;(\si{\mega\ampere})$ & 1.5 & 0.9 \\ \hline
 $B_{tor}\;(\si{\tesla}) $ & 2.1 & 5.4\\ \hline
 $B_{\theta}\;(\si{\tesla})$ & 0.42 & 0.806 \\ \hline
 $n_{OMP}\;(\si{\meter^{-3}})$  & 3.5 $\times 10^{19}$ & 2.11 $\times 10^{20}$ \\ \hline
 $T^{OMP}_{i}\;(\si{\electronvolt})$ & 434 & 165 \\ \hline
 $T^{OMP}_{e}\;(\si{\electronvolt})$ & 160 & 97 \\ \hline
 $n_G\;(\si{\meter^{-3}})$ & 15.6 $\times 10^{19}$ & 7.16 $\times 10^{20}$ \\ \hline
 $\rho_i\;(\si{\meter})$ & 2 $\times 10^{-3}$ & 7 $\times 10^{-4}$ \\ \hline
 $C_s\;(\si[per-mode=fraction]{\meter\per\second})$ & 1.5 $\times 10^{5}$ & 1.07 $\times 10^{5}$ \\ \hline
 $V_{\theta}\;(\si[per-mode=fraction]{\meter\per\second})$ & 4 $\times 10^{4}$ & 1.0 $\times 10^{4}\; \text{to}\; (-1.5) \times 10^{3}$ \\ \hline
 $u_{ti}\;(\si[per-mode=fraction]{\meter\per\second})$ & 1.43 $\times 10^{5}$ & 9.1 $\times 10^{4}$ \\ \hline
 $u_{te}\;(\si[per-mode=fraction]{\meter\per\second})$ & 5.1 $\times 10^{6}$ & 4.5 $\times 10^{6}$ \\ \hline
 $q_{95}$ & 3.7 & 4.0 \\ \hline
 $qR\;(\si{\meter})$ & 8.6 & 3.56 \\ \hline
 $\epsilon$ & 0.34 & 0.294 \\ \hline
 $\omega_{transit}\;(\si{\second}^{-1})$ & 5.9 $\times 10^5$ & 1.26 $\times 10^6$ \\ \hline
 $f_{trapped}$ & 77\% & 72\% \\ \hline
 $\tau_{ii}\;(\si{\second})$ & 5 $\times 10^{-4}$ & 24 $\times 10^{-6}$ \\ \hline
 $\tau_{ei}\;(\si{\second})$ & 3.5 $\times 10^{-6}$ & 1.3 $\times 10^{-7}$ \\ \hline
 $\lambda^{i}_{mfp}\; (\si{\meter})$ & 71.5 & 2.2\\ \hline
 $\lambda^{e}_{mfp}\; (\si{\meter})$ & 17.9 & 0.6\\ \hline
\end{tabular}
\caption{All values are estimated at the outboard midplane.}
\label{table:parameters}
\end{table}

Here, we should make an observation regarding the collisionality between the two discharges: In the case of DIII-D, we have $\epsilon^{3/2}\approx 0.2$, and calculating the dimensionless Coulomb collisionality parametrs $\nu_{\star i} = \frac{\nu_{ii} q_{95} R}{u_{ti}} \approx 0.12$ and $\nu_{\star e} = \frac{\nu_{ei} q_{95} R}{u_{te}} \approx 0.48$, we find that at the edge (OMP), the electrons are in the plateau regime ($\epsilon^{3/2} < \nu_{\star e} < 1$) whereas the ions are in the banana transport regime ($\nu_{\star i} < \epsilon^{3/2}$). For C-Mod though, $\epsilon^{3/2}\approx 0.16$ and $\nu_{\star i} \approx 1.6$, $\nu_{\star e} \approx 6.1$ which places both ions and electrons in the very collisional, Pfirsch-Schl{\"u}tter regime. This is also reflected in the very different mean free paths in the two machines. 

A second observation is related to the ion poloidal flow: Despite the fact that sheared flows are present in both simulations, in the C-Mod case, the ion poloidal flow varies rapidly in a very narrow region near the edge, making it impossible to choose a meaningful single value for $V_{\theta}$. In the DIII-D case the variation is much gentler and the flow doesn't change sign crossing the separatrix.

\section{The Particle and Heat Fluxes}\label{sec:fluxes}
\subsection{Flux definitions}
\begin{figure*}
\centering
\begin{subfigure}[th!]{.5\textwidth}
  \centering
  \includegraphics[width=\linewidth]{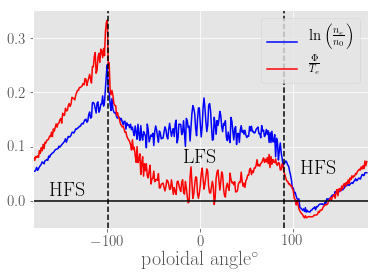}
  \caption{Maxwell-Boltzmann relation in DIII-D.}
  \label{fig:ad_d3d}
\end{subfigure}%
\begin{subfigure}[th!]{.5\textwidth}
  \centering
  \includegraphics[width=\linewidth]{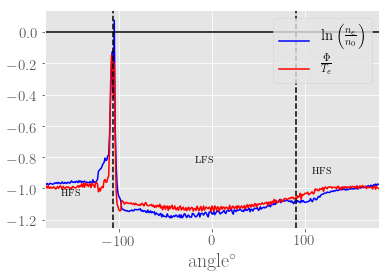}
  \caption{Maxwell-Boltzmann relation in C-Mod.}
  \label{fig:ad_cmod}
\end{subfigure}
\caption{Maxwell-Boltzmann relation in the two machines evaluated on the separatrix. Location of X-points is denoted by dashed lines.\\ {[Associated dataset available at http://dx.doi.org/10.5281/zenodo.3726192]} (\rcite{ioannis_keramidas_charidakos_2020_3726192})}
\label{fig:Ad}

\end{figure*}

In this section we will compare and contrast the particle and heat flux patterns from the two simulations. Our analysis will follow the one in \rcite{keramidas2018analysis} and some figures from there will be repeated here so that comparisons can be made between the edge of DIII-D and the more collisonal edge of C-Mod. First we start with some definitions:
\begin{align}
    n &= \left<n\right>_{t,\zeta} + \delta n \,,\notag\\
    v &= \left<v\right>_{t,\zeta} + \delta v \,,\label{turb_def}
\end{align}

where $n$ and $v$ are the dynamical quantities of plasma density and cross-field velocity and $\left<\cdots\right>$ denotes an average in either time $(t)$ or toroidal planes $(\zeta)$, or both. The time averages considered here are taken over a time interval late in the simulations where a quasi-steady turbulent state has been achieved.

Employing Eq.~\eqref{turb_def}, the fundamental relationship for the fluxes is:
\begin{equation}
    \left<n v\right>_{t,\zeta} = \left<n\right>_{t,\zeta}\left<v\right>_{t,\zeta} + \left<\delta n \delta v\right>_{t,\zeta}\,,\label{flux_def}
\end{equation}
where the cross terms $\left<n \delta v\right>_{t,\zeta}$ and $\left<v \delta n\right>_{t,\zeta}$ vanish due to the vanishing of $\left<\delta n\right>_{t,\zeta}$ and $\left<\delta v\right>_{t,\zeta}$.

As we have pointed out in \rcite{keramidas2018analysis}, these ``fluxes" are not technically transport fluxes but rather local density weighted flows. Because of the very fast electron transit time, the radial excursions of electrons due to classical drifts such as $E \times B$ and magnetic drifts nearly cancel out and contribute negligibly to net transport. Therefore, to obtain a transport flux, one would have top integrate those density weighted flows over a flux surface. Here, as we did before, we will keep referring to them as fluxes for the sake of brevity. We recall that the first piece of the rhs of Eq.~\eqref{flux_def} is known as equilibrium flux, and the second as turbulent flux. (See \rcite{keramidas2018analysis} for additional details.)

The definitions for the heat fluxes are similar, with the replacement of the density by the pressure of each species. Because the simulations are quasineutral, we can not distinguish between the densities or the net particle fluxes of the two species. In the case of the heat fluxes though, a clear distinction between ions and electrons can be made, based on their different temperatures. 

\subsection{Flux Patterns across the separatrix}

\begin{figure*}[t]
\centering
\begin{subfigure}{.5\linewidth}
  \centering
  \includegraphics[height = 6cm, width=\linewidth,keepaspectratio]{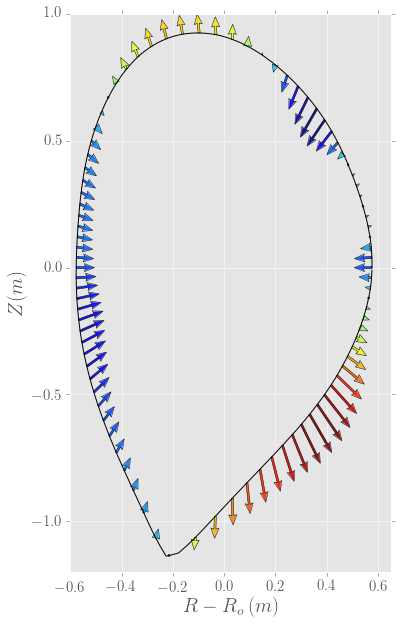}
  \caption{Equilibrium $E\times B$ particle flux in DIII-D.}
  \label{fig:epflux_d3d}
\end{subfigure}%
\begin{subfigure}{.5\linewidth}
  \centering
  \includegraphics[width=\linewidth]{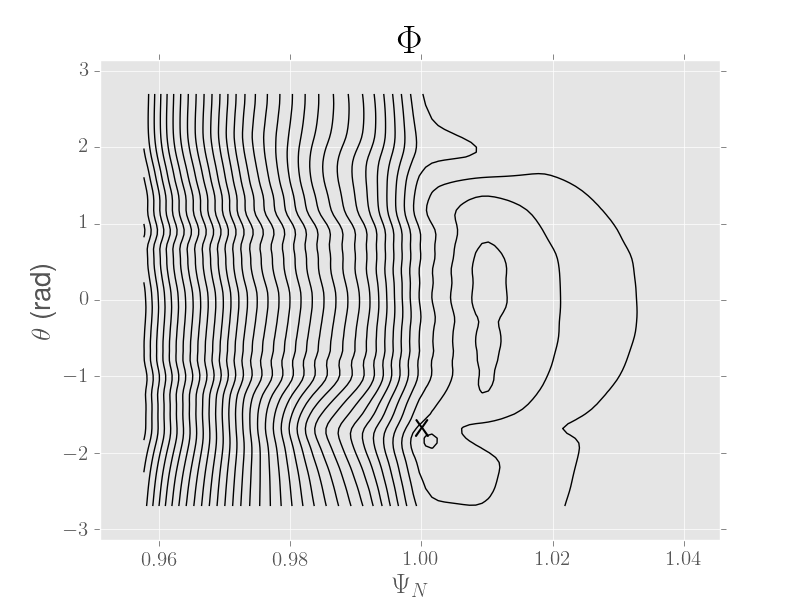}
  \caption{Equipotential contours vs. poloidal angle in DIII-D. \\ Location of lower X-point is indicated with an X-mark.}
  \label{fig:equipot_d3d}
\end{subfigure}\\[1ex]
\begin{subfigure}{.5\linewidth}
  \centering
  \includegraphics[height = 6cm, width=\linewidth,keepaspectratio]{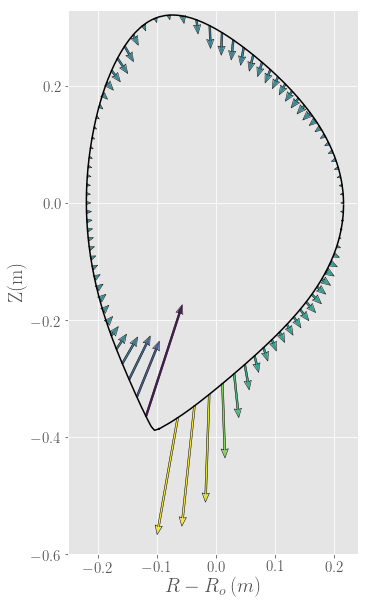}
  \caption{Equilibrium $E\times B$ particle flux in C-Mod.}
  \label{fig:epflux_cmod}
\end{subfigure}%
\begin{subfigure}{.5\linewidth}
  \centering
  \includegraphics[width=\linewidth]{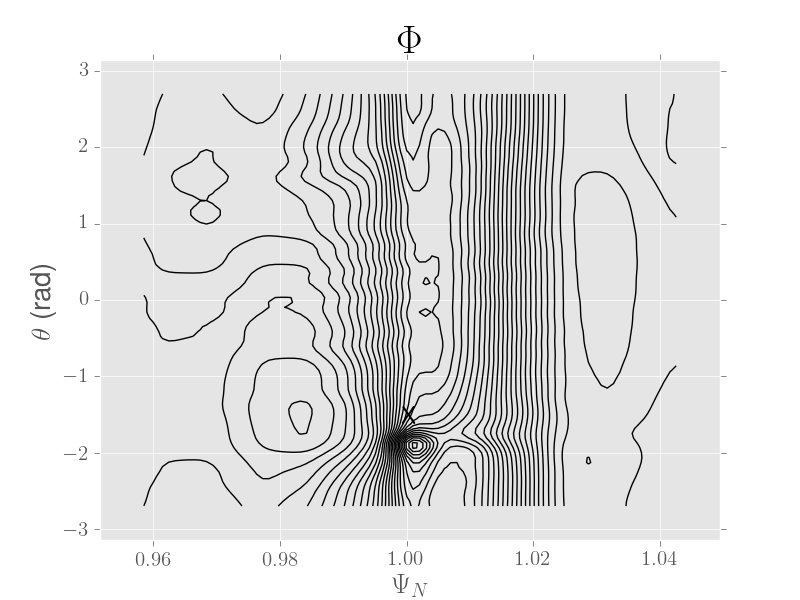}
  \caption{Equipotential contours vs. poloidal angle in C-Mod.\\ Location of lower X-point is indicated with an X-mark.}
  \label{fig:equipot_cmod}
\end{subfigure}
\caption{Equilibrium $E\times B$ particle fluxes and equipotential contours in the two machines. \\ (Strictly, arrow colors correspond to sign and size of local flux but this is easily seen by arrow size and orientation)\\ {[Associated dataset available at http://dx.doi.org/10.5281/zenodo.3726192]} (\rcite{ioannis_keramidas_charidakos_2020_3726192})}
\label{fig:EPF}

\end{figure*}

The first observation regarding the separatrix fluxes has to do with the degree to which the electrons satisfy the Maxwell-Boltzmann (MB) relation at the edge of the two discharges. On one hand, we see (Fig.~\ref{fig:ad_d3d}) that at the LFS midplane of DIII-D there is a significant deviation from MB which we have ascribed to the fact that the transit frequency, the main frequency of the turbulence and the collision rate are all of similar order of magnitude. On the other hand in the C-Mod simulation (Fig.~\ref{fig:ad_cmod}), the MB relation is less modified by any other effects. Indeed, the measured frequency of turbulence in C-Mod, which is close to $\SI{12}{\kilo\hertz}$, is very low compared to the transit frequency of $\SI{1.26}{\mega\hertz}$

Next, in Fig.~\ref{fig:EPF} we present the $E\times B$ particle fluxes around the LCFS. Although the DIII-D case (Fig.~\ref{fig:epflux_d3d}) has been thoroughly analyzed in \rcite{keramidas2018analysis} we will recall here the main points: The inward flux observed at the HFS is due to Pfirsch-Schlutter flows caused by charge polarization from the opposite magnetic drifts of electrons and ions. The interesting alternating pattern of inward and outward flux at the upper and lower LFS respectively, is believed to be due to trapped particle excursions. More specifically, trapped ions that exit from the closed flux surfaces at the bottom, create a charge hole. The local non-adiabaticity means that the electrons can not move instantly along the field lines to neutralize the charge build up. Therefore, a negative potential is created in order to attract more positive charges. The opposite situation takes place at the upper LFS where ions are re-entering. The smaller magnitude of this inward flux compared to the outward one is ascribed to the fact that some ions that leave from the bottom never make it to the top because they get entrained in the parallel flow to the divertors\cite{aydemir2012pfirsch}. 

The situation in C-Mod resembles the one in DIII-D however the equilibrium fluxes are much less prominent. Again, we have an inward flux at the HFS and an alternating inward and outward flux at the LFS with the same polarity as the one in DIII-D. In the C-Mod case though, these features are not as pronounced compared to the very large particle circulation around the X-point. This $E\times B$ circulation has been predicted and explained in \rcite{chang2019x}: the downward magnetic drift of the ions combined with the fact that the poloidal magnetic field at the X-point vanishes logarithmically result in the loss of counter-travelling ions ($u_{\parallel} < 0$) from that point. This ion loss causes a charge and potential build-up in the region. The $E\times B$ flow resulting from this potential has the direction we see in Fig.~\ref{fig:epflux_cmod} and can be intuitively understood as the attempt of the plasma to compensate the weak parallel into-the-divertor flow of counter-travelling ions by forming an into-the-divertor $E\times B$ flow. 

These observations are confirmed from Figs.~\ref{fig:equipot_d3d}, \ref{fig:equipot_cmod} where we draw the equipotential lines as a function of poloidal angle and radial position. The lower X-point is located near $\theta = -1.8$. We observe that in the case of closed field lines near the separatrix, the equipotential lines are straight, indicating that closed flux surfaces share the same potential. The situation however is very different when we move outside the separatrix or close to the X-point. In the case of DIII-D (Fig.~\ref{fig:equipot_d3d}) we find closed potential lines in the LFS. Those closed lines should be interpreted as closed contours of potential hills which is in line with our understanding of charge build up due to trapped ion excursions at these locations. Similarly, in the case of C-Mod (Fig.~\ref{fig:equipot_cmod}), there are closed equipotentials outside the separatrix at the LFS but the most prominent potential hill is at the X-point, and is exactly the positive potentiall hill created by the accumulation of ions that can not be properly dissipated due to the very weak poloidal motion at this point \cite{chang2002x}.

Even though the $E\times B$ circulation around the X-point is expected\cite{chang2019x}, there still remains the issue of why it is so strong in C-Mod but relatively weak in DIII-D. A likely explanation has to do with the collisionality regimes of the two devices: in C-Mod, where the collisionality is very high (cf. Table~\ref{table:parameters}), any potential build up remains strongly localized whereas in DIII-D, which is practically collisionless, a potential perturbation can propagate around the flux surface, influencing the potential around the separatrix in poloidal locations far away from the X-point.

\begin{figure*}
\centering
\begin{subfigure}[t]{.5\linewidth}
  \centering
  \includegraphics[height = 6cm, width=\linewidth,keepaspectratio]{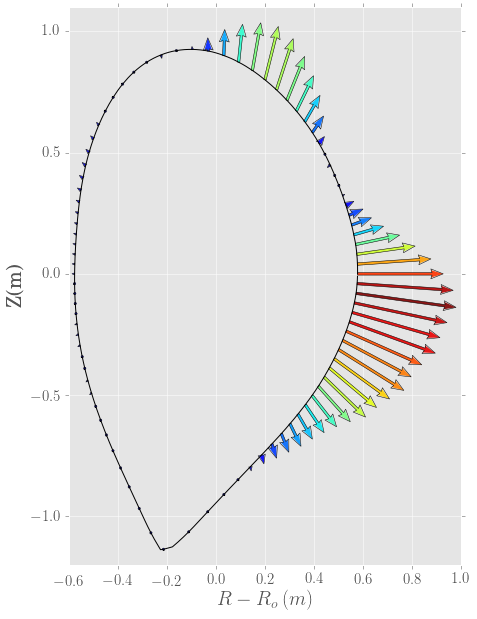}
  \caption{Turbulent $E\times B$ flux vs. poloidal angle in DIII-D.}
  \label{fig:turb_d3d}
\end{subfigure}
\begin{subfigure}[t]{.5\textwidth}
  \centering
  \includegraphics[width=\linewidth]{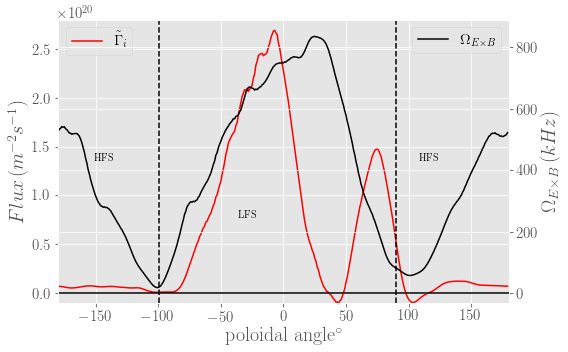}
  \caption{Shear rate vs. turbulence strength in DIII-D.}
  \label{fig:shear_d3d}
\end{subfigure}\\[1ex]
\begin{subfigure}[t]{.5\textwidth}
  \centering
  \includegraphics[height = 6cm, width=\linewidth,keepaspectratio]{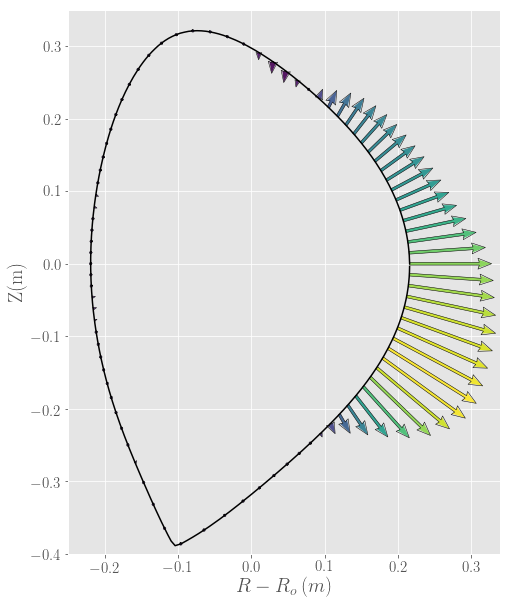}
  \caption{Turbulent $E\times B$ flux vs. poloidal angle in C-Mod.}
  \label{fig:turb_cmod}
\end{subfigure}
\begin{subfigure}[t]{.5\textwidth}
  \centering
  \includegraphics[width=\linewidth]{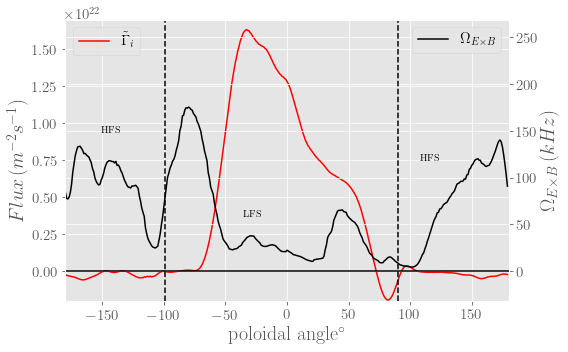}
  \caption{Shear rate vs. turbulence strength in C-Mod.}
  \label{fig:shear_cmod}
\end{subfigure}
\caption{Turbulent fluxes in the two machines. Location of X-points is denoted by dashed lines.\\ {[Associated dataset available at http://dx.doi.org/10.5281/zenodo.3726192]} (\rcite{ioannis_keramidas_charidakos_2020_3726192})}
\label{fig:turbulence}
\end{figure*}

Next, we continue with presenting the poloidal patterns of the turbulent part of the $E\times B$ flux in the two machines (Figs.~\ref{fig:turb_d3d}-\ref{fig:turb_cmod}) and relate them to the respective shear rates (Figs.~\ref{fig:shear_d3d}-\ref{fig:shear_cmod}). In both machines, turbulent $E\times B$ flux is dominant at the LFS. In the case of DIII-D, this flux is interrupted due to the very large shear\cite{burrell1997effects}, as we can infer from Fig.~\ref{fig:shear_d3d}. Indeed, at the region where turbulence is suppressed, the shearing rate is very large and surpasses the main frequency of the turbulence which is close to 600 kHz. In the C-Mod case on the other hand, we find a relatively uniform magnitude of $E\times B$ flux, concentrated at the LFS. A visual inspection of Fig.~\ref{fig:shear_cmod} shows that the shearing rate is apparently not strong enough to influence the magnitude of the turbulent flux. This is in line with our previous comment regarding collisionality: The strong collisionality of C-Mod prevents the potential build up from the X-point loss to have global effects such as enhanced  shear that suppresses the turbulence at the midplane. The collisionless nature of DIII-D on the other hand, allows this to happen. 

\begin{figure*}
\centering
\begin{subfigure}[b]{.5\textwidth}
  \centering
  \includegraphics[width=\linewidth]{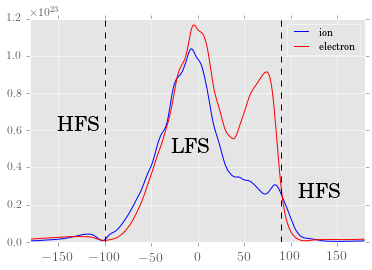}
  \caption{Ion and Electron Turbulent $E\times B$ heat fluxes in DIII-D.}
  \label{fig:heat_d3d}
\end{subfigure}%
\begin{subfigure}[b]{.5\textwidth}
  \centering
  \includegraphics[width=\linewidth]{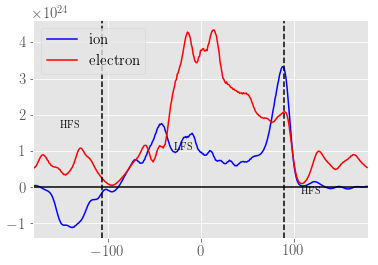}
  \caption{Ion and Electron Turbulent $E\times B$ heat fluxes in C-Mod.}
  \label{fig:heat_cmod}
\end{subfigure}
\caption{Ion and Electron Turbulent $E\times B$ heat fluxes in the two machines. Location of X-points is denoted by dashed lines.\\ {[Associated dataset available at http://dx.doi.org/10.5281/zenodo.3726192]} (\rcite{ioannis_keramidas_charidakos_2020_3726192})}
\label{fig:heat}

\end{figure*}

The last topic in the flux analysis comparison of the two machines concerns the respective heat fluxes. Because of the different temperatures of the two species, here we can make a distinction between electron and ion fluxes. The observation made in \rcite{keramidas2018analysis} was that in DIII-D, the electron turbulent heat flux, shown in Fig.~\ref{fig:heat_d3d}, was significantly larger than the ion one. This was related to the fact that although ions have many different channels for exiting the confinement, the electrons can only do so through turbulence. Therefore, even if the shearing rate is very strong, electron turbulence survives in order to maintain quasineutrality. When we repeat the same analysis in C-Mod, we see that this holds true as well: In Fig.~\ref{fig:heat_cmod}, we find the electron turbulent heat flux to be almost everywhere much larger than the ion one, indicating that this is probably a universal feature of discharges close to the edge. The only physical location where this stops being true is at the lower X-point with the very strong circulation. This is probably related to the pushing of the ion velocity space loss hole to higher energies due to the emergence of the confining electrostatic field\cite{chang2002x}.

\section{Linear Properties of the Simulations}\label{sec:linear}
\begin{figure*}[t]
\centering
\begin{subfigure}{.5\textwidth}
  \centering
  \includegraphics[height = 6cm, width=\linewidth,keepaspectratio]{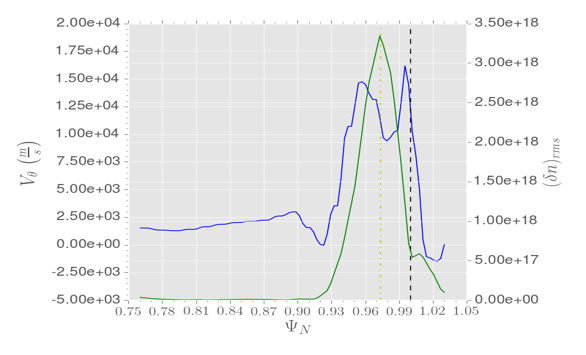}
  \caption{Unstable mode (green) and poloidal flow (blue) in DIII-D. The separatrix is denoted by dashed line.}
  \label{fig:mode_d3d}
\end{subfigure}%
\begin{subfigure}{.5\textwidth}
  \centering
  \includegraphics[height = 6cm, width=\linewidth,keepaspectratio]{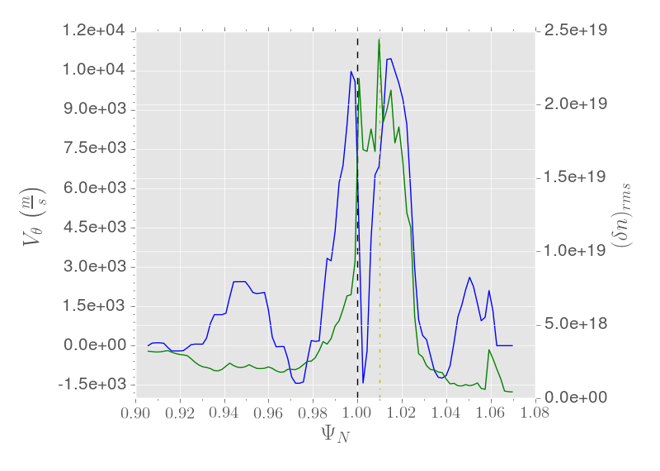}
  \caption{Unstable mode (green) and poloidal flow (blue) in C-Mod. The separatrix is denoted by dashed line.}
  \label{fig:mode_cmod}
\end{subfigure}
\caption{Location of the unstable mode (green) and poloidal flow (black) in the two machines.\\ {[Associated dataset available at http://dx.doi.org/10.5281/zenodo.3726192]} (\rcite{ioannis_keramidas_charidakos_2020_3726192})}
\label{fig:mode}

\end{figure*}

In this section we will describe some properties of the linear modes present in the two simulations. Of course, in a total-f code simulation, like the ones from XGC1, there can be many modes coexisting and interacting with each other. From here on, we will proceed by approximating this complex interaction of modes as a single unstable mode and find its properties by carrying out a Fourier analysis on XGC1 data. In the case of DIII-D we also performed linear electrostatic simulations with the gyrokinetic code {\scshape Gene} assuming that the character of the most unstable linear mode of the {\scshape Gene} simulation would give us some insight into the nature of the unstable mode of the original simulation. As will be seen below, comparison of results from the two methods shows that these assumptions are justified. 

In the case of C-Mod, as evidenced by Fig.~\ref{fig:mode_cmod}, running {\scshape Gene} simulations was not possible because the mode peak (in terms of mode strength $\frac{\delta n}{n_o}$ ) is outside the last closed flux surface where the {\scshape Gene} code is not applicable. As a result, for this simulation we have only the results of the Fourier analysis on the XGC1 output. Doing this revealed a turbulent frequency in the lab frame of about $f_{turb} = \SI{12}{\kilo\hertz}$. The $E\times B$ flow velocity at the mode location (blue line in Fig.~\ref{fig:mode_cmod}) is of the order of $\SI[per-mode=fraction]{3e4}{\meter\per\second}$ which, combined with a poloidal wave number of $k_{\theta} = \SI{314}{\meter^{-1}}$ results in a plasma frame frequency of the order of the Doppler shift and in the electron diamagnetic direction. Because the poloidal $E\times B$ varies rapidly and significantly over the region occupied by the mode (see Table~\ref{table:parameters}), it is not possible to define a precise value for the plasma frame frequency $\omega_{pl}$. The value quoted in Table~\ref{table:parameters2} is an upper estimate. This estimate is larger than but similar to the electron diamagnetic drift frequency $\omega_{\ast e}$; an average over the mode width would reduce $\omega_{pl}$ significantly.  Other than that, we estimated the growth rate to be $\gamma = \SI{3.5e5}{\per\second}$. This estimate comes from fitting an exponential at the initial phase of the simulation and should not be taken to signify the exact linear growth rate of the instability. The start of the simulation is when adjustment of the initial conditions to a non-local neoclassical equilibrium takes place and, unfortunately, there is no way that this process can be separated from the linear instability. Nevertheless, these measurements give us order-of-magnitude estimates about the properties of the linear mode which we will later relate to the features of the turbulent blobs. For now, we remark that the growth rate is roughly $0.1\cdot \omega_{pl}$ in C-Mod with $k_{\perp}\rho_{i}\approx 0.2$, both of which are characteristics of drift waves. Similarly, both are very low compared to the transit frequency and collision rate, something that partly explains the fact that the electron response is very close to Maxwell-Boltzman. 

\begin{table}
\begin{tabular}{|c|c|c|c|}
\hline
\multicolumn{3}{|c|}{Simulation Linear Properties}\\ \hline
    & DIII-D & C-Mod \\ \hline
 $\omega_{lab}\;(\si{\second^{-1}})$ & 4 $\times 10^{6}$ & 7.5 $\times 10^{4}$ \\ \hline
 $\omega_{pl}\;(\si{\second^{-1}})$ & 2.5 $\times 10^{6}$ & 3.14 $\times 10^{6}$ \\ \hline
 $\omega_{\star e}\;(\si{\second^{-1}})$ & 1.96 $\times 10^{6}$ & 2.17 $\times 10^{6}$ \\ \hline
 $\gamma_{G\!E\!N\!E}\;(\si{\second^{-1}})$ & 2.54 $\times 10^{5}$ & ----- \\ \hline
 $\gamma_{L\!P}\;(\si{\second^{-1}}) $ & 1.74 $\times 10^{5}$ & 3.5 $\times 10^{5}$\\ \hline
 $\gamma_{M\!H\!D}\;(\si{\second^{-1}})$ & 1.13 $\times 10^{6}$ & 7.8 $\times 10^{5}$ \\ \hline
 $k_{\perp}\;(\si{\meter^{-1}})$  & 123 & 314 \\ \hline
 $k_{\perp}\rho_{i}$ & 0.246 & 0.22 \\ \hline
 $\Omega_{E\times B}\;(\si{\second^{-1}})$ & 3.8 $\times 10^{6}$ & 1.25 $\times 10^{5}$ \\ \hline
\end{tabular}
\caption{All values are estimated at the outboard midplane.}
\label{table:parameters2}
\end{table}

As we mentioned above, for the DIII-D case we have results from linear, electrostatic runs of {\scshape Gene}. In Fig.~\ref{fig:mode_d3d}, we present the shape of the unstable mode in the $R-Z$ plane, where we have plotted the mode strength $\left( \delta n\right)_{rms}$ at the outboard midplane during the quasi-steady turbulent phase of the simulation. There, we see that the mode peaks inside the separatrix therefore, we can run {\scshape Gene} at this location ($\Psi_N = 0.97$) and scan a range of parameters. All {\scshape Gene} simulations were local, with the center of the box being the mode peak location. The local parameters of density and electron and ion temperature were used (cf. Table.~\ref{table:parameters}). Because the profiles vary widely across the mode, the scale lengths used for the simulation were an average of the scale lengths across the mode, weighted by the mode strength, i.e., $\displaystyle \frac{1}{L_x} = \frac{\int dR\; \left(\delta n\right)^2 \nabla x}{\int dR\; \left(\delta n\right)^2 x}\;$, where $x = n, T_i, T_e$ and $\left(\delta n\right)^2$ is the mode strength. Using the numbers $\frac{1}{L_n} = \SI{42.45}{\meter^{-1}}$,$\frac{1}{L_{T_i}} = \SI{13.82}{\meter^{-1}}$, $\frac{1}{L_{T_e}} = \SI{37.41}{\meter^{-1}}$, we employed the following scheme for linear runs: We labelled the case with the above parameters as our `base' case and we carried out all different combinations of varying the inverse scale lengths between half and double their base value. Therefore, we ended up with 27 different runs that demonstrate the response of the linear modes of the system to the variation of the different drivers. The geometric information was specified by the same EFIT file used in the original XGC1 simulation and all {\scshape Gene} runs were confined to wavenumbers up to $k_{\perp}\rho_i = 2.0\;$ which is the finest scale that the XGC1 mesh can distinguish. 

\begin{figure*}
\centering
\begin{subfigure}[t]{.5\textwidth}
  \centering
  \includegraphics[height = 6cm, width=\linewidth]{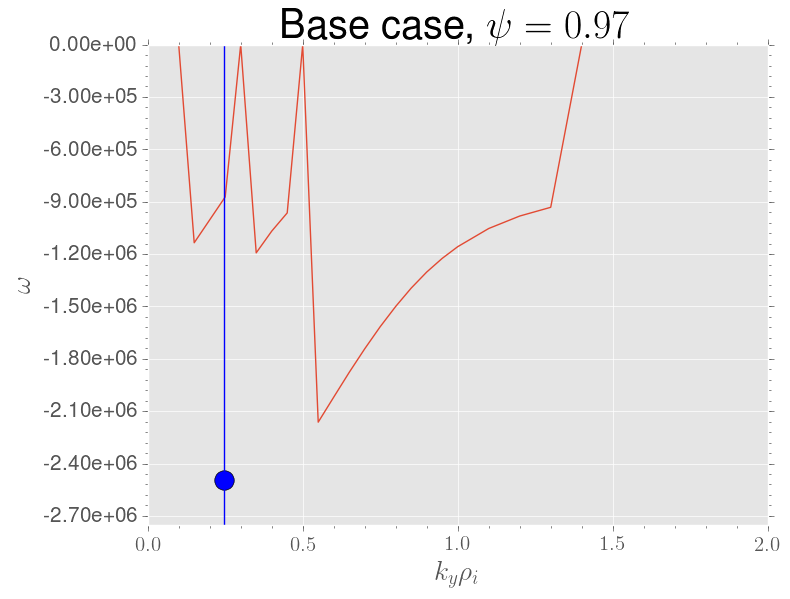}
\caption{}
  \label{fig:gene_base_freq}
\end{subfigure}%
\begin{subfigure}[t]{.5\textwidth}
  \centering
  \includegraphics[height = 6cm, width=\linewidth]{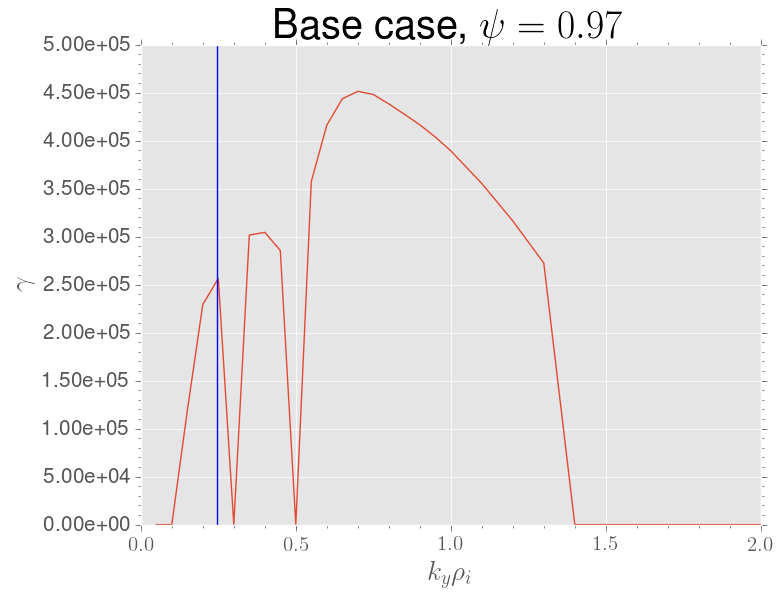}
  \caption{}
  \label{fig:gene_base_growth}
\end{subfigure}
\caption{(a) Real frequency (b) Growth rate of {\scshape Gene} base case run of DIII-D. In both panels, the XGC1 measured wavenumber is represented by the vertical blue line. In panel (a) the XGC1 measured frequency is represented by a blue dot.\\ {[Associated dataset available at http://dx.doi.org/10.5281/zenodo.3726192]} (\rcite{ioannis_keramidas_charidakos_2020_3726192})}
\label{fig:gene_base}

\end{figure*}

In Fig.~\ref{fig:gene_base} we present the real frequency and linear growth rate of the base case {\scshape Gene} run. Out of the three unstable modes that {\scshape Gene} predicts, the first one has a growth rate that peaks exactly at the wavenumber that we find from the XGC1 data and has a linear frequency in the same direction (electron diamagnetic) as the one measured in the simulation (Doppler shifted back into the plasma frame). Estimating the growth rate from the XGC1 data by fitting an exponential at the linear phase of the simulation, we find a growth rate of $\gamma = \SI{1.74e5}{\second^{-1}}$. With the reminder that this number is contaminated by the adjustment of the equilibrium, the {\scshape Gene} result of $\gamma = \SI{2.54e5}{\second^{-1}}$ is judged to be in satisfactory agreement. 

\begin{figure}
    \centering
    \includegraphics[height = 8cm, width=\linewidth]{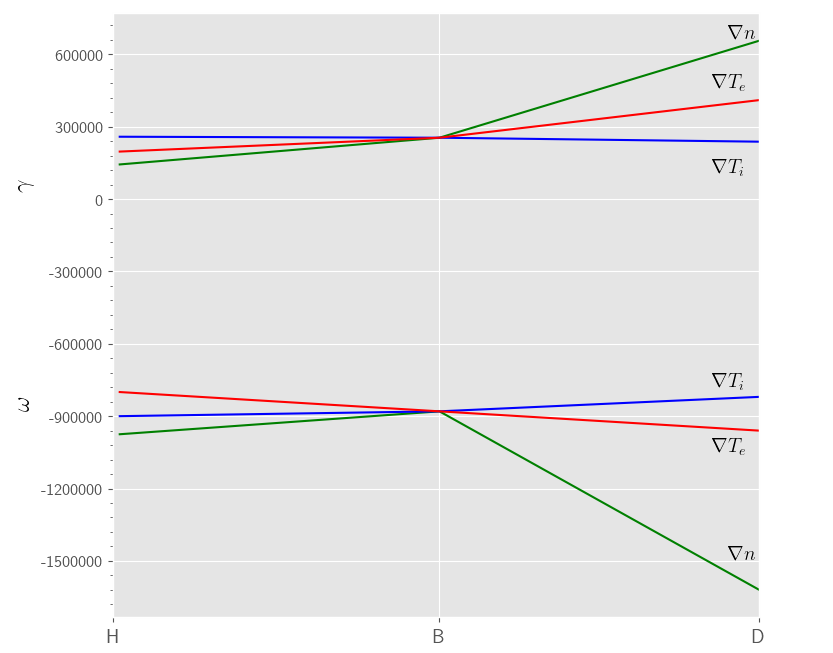}
    \caption{Response of the linear growth rate and real frequency of the unstable mode to the variation of the turbulent drivers.\\ {[Associated dataset available at http://dx.doi.org/10.5281/zenodo.3726192]} (\rcite{ioannis_keramidas_charidakos_2020_3726192})}
    \label{fig:drivers}
\end{figure}

The response of the linear growth rate and the real frequency of the unstable mode to the variation of the three turbulent drivers, $\nabla n$, $\nabla T_i$, $\nabla T_e$, between half and double their base value can be seen in Fig.~\ref{fig:drivers}. Each of the three lines represents three cases where the highlighted driver is varied between its extreme values and the other two scale lengths are kept at their base values. It is evident from the trends that the basic drivers of the mode are the density and electron temperature gradients and that the ion temperature gradient increase seems to have a mild stabilizing effect. We can attribute the later to the influence of finite Larmor radius (FLR) effects while the destabilizing effect of $\nabla n$ and $\nabla T_e$, along with the direction of propagation ($\omega_{\star e}$) and the ion-scale of the instability,  point us to a Trapped Electron Mode. It is worth mentioning that we have also estimated the ideal MHD growth rate, $\gamma_{M\!H\!D} = \sqrt{\frac{2 c^{2}_s}{R L_p}}$, where $L_p$ is the scale length of the pressure. We found that $\gamma_{M\!H\!D} = \SI{1.13e6}{\second^{-1}}$ which is of the same order of magnitude as the XGC1 measured turbulent frequency and larger than the {\scshape Gene} and XGC1 growth rates. This is an indication that interchange modifications must have a strong influence on the linear mode\cite{guzdar1993three}. In Table~\ref{table:parameters2} we have gathered all the calculated numbers of the linear properties of the modes to aid the reader.

\begin{figure*}[t]
\centering
\includegraphics[height=8cm, width=\linewidth]{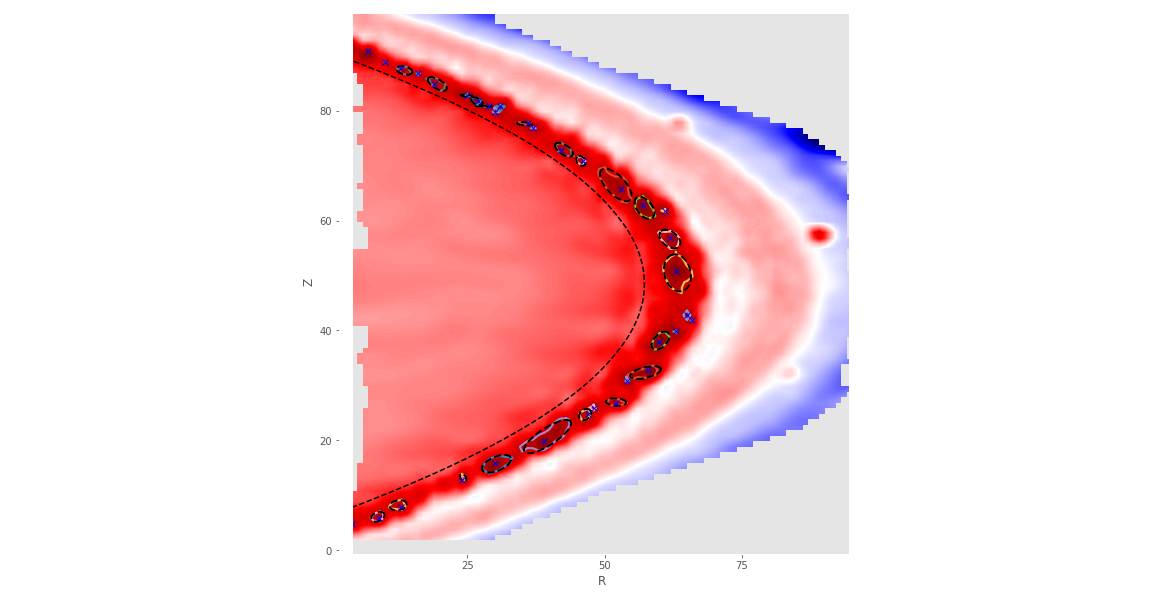}
\caption{Density perturbation color plot of a frame from the C-Mod outboard midplane. We can see the identified blob contours in colored lines and the fitted ellipses in black dashed lines. Blob peaks are denoted with $\times$ symbols. The separatrix is the curved black dashed line. The units of the axes are in pixels. All of the blobs are in the SOL.}
\label{fig:blob_module}
\end{figure*}

\section{Properties of Blob Filaments}\label{sec:blobs}
\subsection{Blob Detection and tracking}

The XGC1 simulations produce a large number of blob filaments, originating close to the edge and propagating into the SOL. In this work, we want to study the properties of these structures in relation to the unstable modes. We therefore developed a blob detection and tracking module in Python for use with the XGC1 data. 

The method we used follows roughly in the steps of \rcite{zweben2016blob}. Before we feed the data into the blob detection algorithm, we submit them to a two-stage preprossesing routine: First, we perform  a two-dimensional smoothing over the range of each frame in the $R-Z$ plane, where we replace the value of each pixel with a weighted average of it's nearest neighbors. After that, we also smooth along the field lines, since blob filaments are field-aligned structures, replacing the value of each pixel with the weighted average of it's toroidal neighbouring points that lie on the same field line. 

The detection algorithm uses an off-the-shelf Python routine for locating contours above a certain threshold. The distinguishing characteristic that we require from a structure in order to qualify as a blob is that the density perturbation $\frac{\delta n}{n}$ exceeds a certain threshold above the background, which here, we took (arbitrarily) to be $20\%$. After the Python routine tracks all contours above said threshold, each blob is defined by its peak, that is, given that a set of contours contains the same maximum, we reject all but the largest one. In this way, we avoid double counting as, in the end, each blob contour contains a single, unique peak. Finally, an ellipse is fitted to each blob contour using the algorithm of \rcite{fitzgibbon1999direct} and a list of parameters such as blob area, peak value and location, length of major and minor axis of the ellipse, tilt angle etc. are stored in an SQL database for easy retrieval. To give an idea about how the detection module works, we provide Fig.~\ref{fig:blob_module} where we see a density perturbation color plot of a time frame from the C-Mod outboard midplane. The blob contours are demarcated by colored lines which enclose the blob peaks (illustrated by $\times$ symbols) and on top of them, we observe the black dashed lines of the fitted ellipse. The few peaks that have no enclosing contours or no fitted ellipses around them are either too small or too close to the edge of the frame so that we can not find a closed contour. Those blobs are not included in the database and subsequent analysis. 

The tracking feature of the module is implemented as follows: for each simulation time step, we take all blobs present at that time and use a comparison algorithm to compare them to each blob present in the following time step. The tracking algorithm computes a score based on the fact that  the same blobs would have a radial and poloidal velocity that would follow a roughly normal distribution with means and variances  chosen from experimental considerations and that their areas, like their velocities,  should also not change dramatically between frames. Using this procedure, with the distribution functions properly calibrated, generally results in an unambiguous $1\!-\!1$ matching between blobs from adjacent time frames. Occasionally, we are able to track the splitting of one blob into two or more, or the inverse process of blob mergers. In this paper, we will not be presenting results from the tracking feature of the module. 

\begin{figure*}[t]
\centering
\begin{subfigure}{.5\textwidth}
  \centering
  \includegraphics[height = 6cm, width=\linewidth,keepaspectratio]{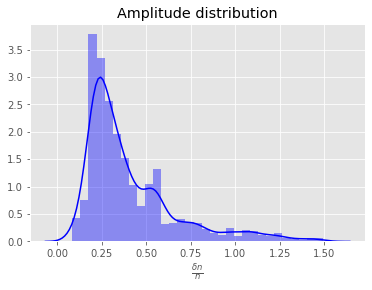}
  \caption{Amplitude distribution of blobs in DIII-D.}
  \label{fig:d3d_amplitude}
\end{subfigure}%
\begin{subfigure}{.5\textwidth}
  \centering
  \includegraphics[height = 6cm, width=\linewidth,keepaspectratio]{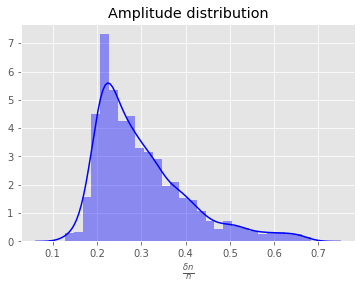}
  \caption{Amplitude distribution of blobs in C-Mod.}
  \label{fig:cmod_amplitude}
\end{subfigure}
\caption{Amplitude distributions of blobs in the two machines.\\ {[Associated dataset available at http://dx.doi.org/10.5281/zenodo.3726192]} (\rcite{ioannis_keramidas_charidakos_2020_3726192})}
\label{fig:blob_ampl}
\end{figure*}

\begin{figure*}[t]
\centering
\begin{subfigure}{.5\textwidth}
 \centering
  \includegraphics[height = 6cm, width=\linewidth,keepaspectratio]{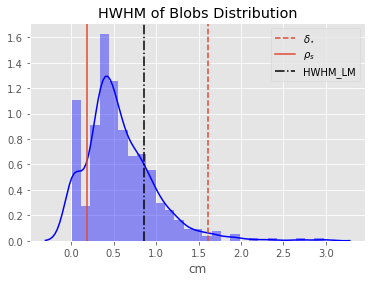}
  \caption{Size distribution of blobs in DIII-D.}
  \label{fig:d3d_size}
\end{subfigure}%
\begin{subfigure}{.5\textwidth}
  \centering
  \includegraphics[height = 6cm, width=\linewidth,keepaspectratio]{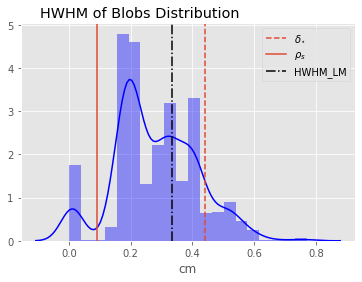}
  \caption{Size distribution of blobs in C-Mod.}
  \label{fig:cmod_size}
\end{subfigure}
\caption{Size distributions of blobs in the two machines.\\ {[Associated dataset available at http://dx.doi.org/10.5281/zenodo.3726192]} (\rcite{ioannis_keramidas_charidakos_2020_3726192})}
\label{fig:blob_size}
\end{figure*}

\subsection{Blob features}

Using the blob detection module, filaments are identified in both simulations after the linear phase. In the case of DIII-D, the total number of blobs that fulfill the criteria we set in order to include them in the statistics are 741 whereas in the case of C-Mod, 1332. We focus on two static blob properties (we will present statistics from blob dynamics in the future), namely the amplitude and the size. The blob amplitude is compared to what has been experimentally observed and the blob size set against other characteristic sizes of the problem and we draw conclusions about the nature of the generated blobs.  

In Fig.~\ref{fig:blob_ampl} we give the probability distribution functions for the blob amplitudes in the two simulations. Both plots show the fitted Gaussian kernel density in blue (the y-axis units are in units of probability density). We see that blobs from both simulations have an amplitude that is roughly exponentially distributed, a fact that has been observed experimentally in MAST \cite{militello2018two}. 

The blob size is shown in Fig.~\ref{fig:blob_size}. To define a size for the blob we have experimented with various choices for a blob profile (shape). We found that the most reasonable choice for our data was to assume that the blob profile corresponds to the positive part of a sinusoid that has been convected out to the measurement location. (Recall that nonlinearly curvature-interchange dynamics propels positive fluctuations radially outward and negative fluctuations inwards.) Then, by ``blob size'' we define the half width at half maximum (HWHM) of this sinusoid. In Appendix.~\ref{app}, we discuss the procedure by which we reconstruct this size of each blob from the data we have recorded in the database. 

In Fig.~\ref{fig:blob_size} the histogram of the blob sizes is compared with the Larmor radius $\rho_s$, the dimensional characteristic blob scale size $\delta_{\ast}$ as defined in \rcite{myra2006collisionality}, $\delta_{\ast} = \rho_s\left(\frac{L^2_\parallel}{\rho_s R}\right)^{\frac{1}{5}}$, with $L_\parallel$ being the connection length and $R$ the major radius, and the HWHM of the linear mode (HWHM$_{LM}$), defined by $\text{HWHM} = \frac{\pi}{3k_\perp}$, with $k_\perp$ taken from Table~\ref{table:parameters2}. We note that even though the shape of these histograms is insensitive to the assumed form of the blob profile, different profile choices can change how the histogram maps to the x-axis. Nevertheless, whichever blob profile is postulated, the peak of the blob size distribution was found to have a scale smaller than that of the linear mode (HWHM$_{LM}$), which may be consistent with sheared flows tearing up the linear structures. Also worth noting is the fact that there are very few blobs above the $\delta_{\ast}$ scale, which is consistent with the fact that none of the observed structures connects to the divertor plate sheaths. Recall here that in the vorticity charge conservation equation, blobs with $\delta < \delta_{\ast}$ are dominated by inertial (ion polarization drift) currents while blobs with $\delta > \delta_{\ast}$ are dominated by parallel current flow to the sheath\cite{myra2006collisionality, Krasheninnikov2008review}. This is to be expected since XGC1 effectively cuts off currents into the sheath implementing a logical sheath boundary condition which modifies the sheath potential trying to enforce ambipolar fluxes to the wall\cite{churchill2017kinetic}. 

\section{Summary and Conclusions}\label{sec:concl}
We have presented and compared results from the analysis of two XGC1 simulations. Regarding the separatrix fluxes, both DIII-D and C-Mod simulations revealed a similar equilibrium $E\times B$ poloidal flux pattern. This pattern has previously been ascribed, in the DIII-D case, to a combination of $\nabla B$-drifts and trapped ions exiting and re-entering the closed surface region. We corroborated this conclusion by providing the equipotential contour plots in the $\Psi-\theta$ plane which reveal potential structures localized at the LCFS that resemble charge accumulations. We showed that this pattern, in principle, holds in C-Mod as well albeit, much less attenuated compared to the very strong X-point circulation. The reason for the X-point circulation is the well known X-point loss however, we have attributed the big difference in the strength of it between the two machines, to the rather large difference in their collisionalities. The high collisionality of C-Mod forces the X-point loss potential build-up to be localized whereas, in the practically collisionless DIII-D case, any potential perturbation will travel around the flux surface and influence the potential at remote poloidal locations. This line of reasoning can also explain why in DIII-D the shear reaches (locally) high enough levels in order to suppress the turbulent flux but in the C-Mod case, it may remain too low to influence the turbulence. As far as the heat fluxes are concerned, we have shown that in both simulations the electron turbulent heat flux at the edge is  larger than the ion one, which is probably a universal feature of discharges. This is due to the very small Larmor radius of the electrons which leaves them only turbulence as a means by which to exit the confinement and maintain quasineutrality.  

For both simulations, Fourier analysis has revealed turbulent frequencies and an approximation of the growth rates of the unstable modes. In the case of DIII-D, we also did a linear analysis using the {\scshape Gene} code. The linear analysis was found to be in close agreement with the measured frequencies and growth rates. The parameter scan that we did in order to find the response of the most unstable mode to the turbulence drivers, along with the frequency direction and spatial scale of the instability, indicate the presence of a TEM mode with strong interchange modifications. For the C-Mod case, because the mode peaks outside the LCFS, we could not perform a {\scshape Gene} simulation. However, the measured frequency, growth rate and spatial scale of the mode, all suggest a drift wave.

We also presented the basic features of a blob detection and tracking module that we created for the XGC1 data. The module has the ability to identify individual blobs, fit an ellipse around them and store relevant information about the filament in a database. It can also track blobs from one time frame to another, measuring their velocities. Here, we focused on two static features of the blobs, the amplitude and size distributions. The distribution of blob amplitudes was found to be roughly exponential, in qualitative agreement with previous experimental observations. The distribution of blob sizes reveals that most of them cluster between the Larmor radius and the size of the unstable mode. This is expected from structures that are created by the shearing of the turbulent mode. Moreover, almost all of the blobs are smaller than the characteristic size $\delta_{\ast}$ that is relevant for blobs that connect to the divertor and are influenced by current to the sheath. Therefore, we deduce that in both simulations, blob filaments are dominated by ion polarization currents. In future publications we would like to explore the dynamic and static blob properties from these and other XGC1 simulations and draw conclusions about the connection between edge flows, edge fluxes, turbulence and blob properties.  

\section*{Acknowledgements}
This work was supported by the U.S. Department of Energy Office of Science, Office of Fusion Energy Sciences under grant DE-FG02-08ER54954 and DE-SC000801 and by subcontract SO15882-C with PPPL under the U.S. Department of Energy HBPS SciDAC project DE-AC02-09CH11466. The DIII-D US DOE Grant number for the particular shot that was used as a basis for the analysis was DE-FC02-04ER54698 and for the C-Mod one, DE-FC02-99ER54512. The original XGC1 runs used computing resources on Titan at OLCF through the 2015 Innovative and Novel Computational Impact on Theory and Experiment (INCITE) program and the 2016 ALCC (ASCR Leadership Computer Challenge) award. 

\section*{Data Availability}
Digital data for all the figures of this paper (except Fig.~\ref{fig:blob_module} and Fig.~\ref{fig:blob_draw}) are openly available in Zenodo, at https://doi.org/10.5281/zenodo.3726192 with DOI 10.5281/zenodo.3726192, \rcite{ioannis_keramidas_charidakos_2020_3726192}.

\appendix
\section{Blob Size Extracted From the Data}\label{app}

\begin{figure}
    \centering
    \includegraphics[width=\linewidth]{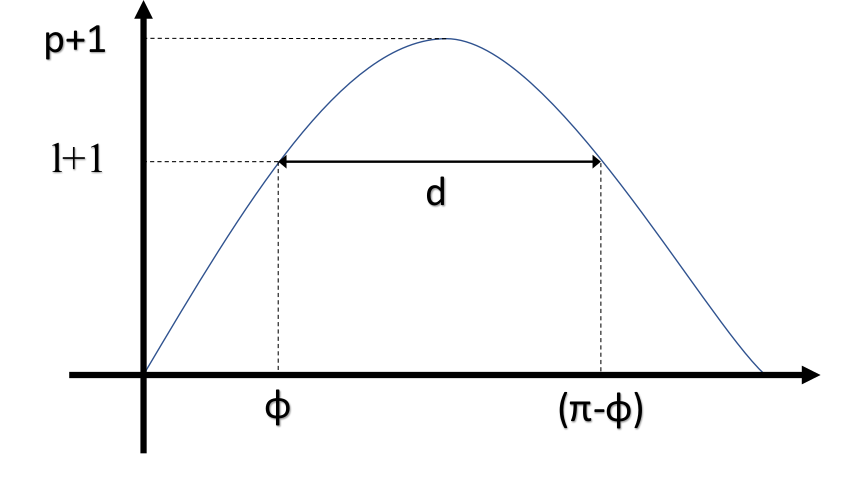}
    \caption{Illustration of the blob features.}
    \label{fig:blob_draw}
\end{figure}

As we mentioned above, the blobs are assumed to have the shape of the positive part of a sinusoid. The blob detecting module records in the database the peak, $p$, and level, $l$, values of the blob. By level value, we mean the $\frac{\delta n}{n_o}$ of the base contour of the blob, i.e., the largest contour that can still be considered part of the blob. In Fig.~\ref{fig:blob_draw} we illustrate the situation, assuming that the blob is the positive part of the sinusoid oscillation $n = n_p \sin(kx) = (p+1) \sin(kx)$. Assuming that we know the size $d$ at the level contour, then $d$ is related to the wavenumber of the sine wave by $d = k (\pi - 2\phi)$\,, with $\phi$ being the phase of the oscillation where we find the level contour. From the definition of the sine wave, this phase is given by $\phi = \arcsin\left(\frac{l+1}{p+1}\right)$. Combining the previous two equations, we get a relationship between $d$ and $k$. From that, we can arrive at the equation for the HWHM  ($h=\frac{\lambda}{6}$), 
\begin{equation}\label{eq:hwhm}
h = \frac{\pi d}{3\left(\pi - 2 \arcsin\left(\frac{l+1}{p+1}\right)\right)}\,.
\end{equation}

Instead of the size $d$, at the database we have stored the major radius $R_M$ of the fitted ellipse. To find $d$, we need to project this length into the binormal direction. This direction is taken to be $\hat{e}_\chi = \hat{b}\times \hat{e}_\psi$\,, which, after a little manipulation, results in the explicit formula $\hat{e}_\chi = \frac{1}{|B|B_p} \left(-B_\zeta B_R \hat{e}_R -B_\zeta B_Z \hat{e}_Z + (B^2_R +B^2_Z) \hat{e}_\zeta\right)\,,$ with $|B| = \sqrt{B^2_R + B^2_Z + B^2_\zeta}$ and $B_p = \sqrt{B^2_R + B^2_Z}\,$.
The projected-to-the-binormal blob size is then:
\begin{equation}\label{eq:bn}
    d = \frac{1}{|B|B_p} \left(-B_\zeta B_R d_R -B_\zeta B_Z d_Z \right)\,.
\end{equation}
where,
\begin{align*}
d_R =& \cos(\theta) R_M\,,\\ 
d_Z =& \sin(\theta) R_M\,.
\end{align*}
with $\theta$ being the blobs' tilt angle. To be more precise, $d$ is the binormal projection of the blob size on the R-Z plane. Plugging Eq.~\ref{eq:bn} into Eq.~\ref{eq:hwhm}, we find the size of each blob in the database.

\bibliographystyle{unsrt}
\bibliography{fluxbibliography}{}

\end{document}